\documentclass[12pt]{article}
\usepackage{graphicx,psfrag,epsfig}
\usepackage{amssymb,amsmath,amscd,amsthm}
\usepackage{graphicx,psfrag,epsfig}

\usepackage{graphicx}
\usepackage[active]{srcltx}
\usepackage{epsfig,color}

\newtheorem{theorem}{Theorem}[section]

\newtheorem{lemma}[theorem]{Lemma}

\begin{document}
\title{On mathematical foundation of the Brownian motor theory.}
\author{ L. Koralov\footnote{Dept. of Mathematics, University of Maryland,
College Park, MD 20742, koralov@math.umd.edu}, S.
Molchanov\footnote{Dept of Mathematics and Statistics, University
of North Carolina, Charlotte, NC 28223, smolchan@uncc.edu;
corresponding author, phone 704-6874573}, B.
Vainberg\footnote{Dept of Mathematics and Statistics, University
of North Carolina, Charlotte, NC 28223, brvainbe@uncc.edu}}
\date{}

\maketitle

\begin{abstract}
The paper contains mathematical justification of basic facts
concerning the Brownian motor theory. The homogenization theorems
are proved for the Brownian motion in periodic tubes with a
constant drift. The study is based on an application of the Bloch
decomposition. The effective drift and effective diffusivity are
expressed in terms of the principal eigenvalue of the Bloch
spectral problem on the cell of periodicity as well as in terms of
the harmonic coordinate and the density of the invariant measure.
We apply the formulas for the effective parameters to study the
motion in periodic tubes with nearly separated dead zones.
\end{abstract}



\section{Introduction}

The paper is devoted to mathematical theory of Brownian
(molecular) motors. The concept of a Brownian motor has
fundamental applications in the study of transport processes in
living cells and (in a slightly different form) in the porous
media theory. There are thousands of publications in the area of
Brownian motors in the applied literature. For example, the review
by P. Riemann \cite{ rei} contains 729 references. Most of these
publications are in physics or biology journals and are not
mathematically rigorous. Some of them are based on numerical
computations.

Consider a set of particles with an electrical charge performing
Brownian motion in a tube $\Omega$ with periodic (or stationary
random) cross section (see Fig. 1).
\begin{figure}[tbph]
\begin{center}
 \includegraphics[width=0.75\columnwidth]{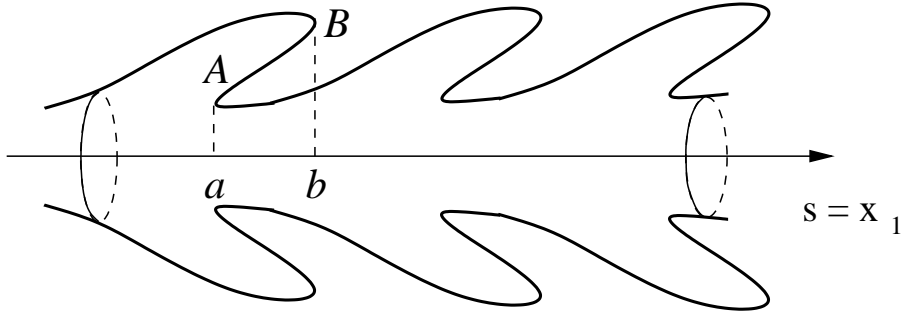}
\end{center}
\caption{A periodic tube $\Omega$ with ``fingers". One can expect
that $V_{\rm eff} <|V_{\rm eff}^{-}|$.} \label{extj2}
\end{figure}

We will assume that the axis of $\Omega$ is directed along the
$x_1$-axis. Let's apply a constant external electric field $E$
along the axis of $\Omega$. Then the motion of the particles
consists of the diffusion and the Stocks drift, and the
corresponding generator has the form
\[
L u = \Delta u +V \frac{\partial u }{\partial x_1},~~~V=V(E),
\]
 complemented by the Neumann boundary condition on
$\partial\Omega$ (we assume the normal reflection at the
boundary).

One can expect that the displacement of the particles on the large
time scale may be approximated (due to homogenization) by a
one-dimensional diffusion process $\overline{x}_1(t)$ along the
$x_1$-axis with an effective drift $V_{\rm eff}$ and an effective
diffusivity $\sigma^2$, which depend on $V$ and the geometry of
the tube $\Omega$. If we reverse the direction of the external
field from $E$ to $-E$, then the corresponding effective drift
$V_{\rm eff}^-$ and the corresponding effective diffusivity
$(\sigma^-)^2$ will be, generally, different from $V_{\rm eff}$
and $\sigma^2$ when $\Omega$ is not symmetric with respect to the
reflection $x_1\to -x_1$. For example, one can expect that $V_{\rm
eff}<|V_{\rm eff}^-|$ for $\Omega$ shown in Fig.~1. The difference
between the effective parameters can be significant. Then by
changing the direction of the exterior electric field $E$
periodically in time, one could construct a constant drift in,
say, the negative direction of $x_1$ and create a device (motor)
producing energy from Brownian motion.

The idea of molecular motors goes back to M. Smoluchowski, R.
Feynman, L. Brillouin. Starting from the 1990-s, it became a hot
topic in chemical physics, molecular biology, and thermodynamics.
The following natural problem must be solved by mathematicians:

1. The homogenization procedure (reduction to a one-dimensional
problem) must be justified.

2. Expressions for effective parameters $V_{\rm eff}, \sigma^2$ in
terms of an appropriate PDE or a spectral problem on the period of
$\Omega$ (in the case of periodic tubes) must be found.

3. For some natural geometries of $\Omega$ that include a small
parameter, asymptotic expressions for effective parameters need to
be obtained.

We solve the  first two problems here using an analytic approach
that justifies the homogenization procedure and allows us to
express the effective parameters in terms of the principal
eigenvalue of a spectral problem on one period of $\Omega$. We
will also provide a couple of simple consequences of the obtained
formulas. We show that ${d{V_{\rm eff}}}/{d V}=\sigma^2>0$ when
$V=0$. Thus, since $V_{\rm eff}$ is analytic in $V$, it is a
strictly monotone function of $V$ when $|V|$ is small enough.
Simple asymptotic formulas for the effective parameters will be
justified for periodic tubes with nearly separated dead zones.
Note that the first two problems listed above also can be solved
with probabilistic techniques similar to those used for
homogenization of periodic operators in $\mathbb{R}^d$, as we'll
discuss in a forthcoming paper.

Our approach is based on the Bloch decomposition. Recently several
papers appeared (see \cite{CoVann,CoOrVann} and references there),
where the Bloch decomposition was used to study homogenization
problems in periodic media. It was shown that this approach has
many advantages (when it is applicable). So far, this approach was
applied to self-adjoint  elliptic equations in the whole space or
in a domain with a finite boundary. The symmetry and the existence
of a bounded inverse operator were essential there.

We consider a parabolic problem. It is non-self-adjoint due to the
drift, which can not be neglected since it is essential for
applications. Besides, the homogenization is applied only with
respect to one variable. All these features (the non-symmetry is
the main difficulty) make the problem under consideration
essentially different from the applications of the Bloch
decomposition in the homogenization mentioned above.

We also would like to note the presence of the boundary integral
term in the formula for the effective drift (see Theorem
\ref{gt}). In probabilistic terms, it arises from the
re-normalized time spent by the Brownian motion with the drift on
the boundary (local time). Our next paper based on a probabilistic
approach to the problem will contain the detailed analysis of the
asymptotic behavior of the effective parameters with respect to
the parameters describing the geometry of the domain.

The plan of the paper is as follows. Lemma \ref{ml} describes the
properties of the principal eigenvalue
$\lambda_0=\lambda_0(\theta)$ of the Bloch spectral problem on the
cell of periodicity of $\Omega$. The main results on
homogenization are obtained in Theorems \ref{t1} and \ref{t2}. In
particular, it is shown there that  the effective drift $V_{\rm
eff}$ and the effective diffusivity $\sigma^2$ are given by the
coefficients of the Taylor expansion of the eigenvalue
$\lambda_0(\theta)$. To be more exact, $\lambda_0(\theta)=iV_{\rm
eff}\theta-\sigma^2\theta^2+O|\theta|^3,~\theta\to 0.$
Expressions for the effective parameters through the harmonic
coordinate and the invariant measure are given in Theorem
\ref{gt}. The latter formulas are applied to a particular class of
domains in the last section. Periodic tubes with nearly separated
dead zones are considered there. We rigorously justified the
asymptotic formula for the effective diffusivity, which was found
earlier for domains with somewhat simpler geometry in \cite{bm1},
\cite{bm}.

\section{Description of the model and main results.}

Consider a tube $\Omega \subset \mathbb{R}^{d}$, $d\geq 2,$
periodic in $x_1$ with period 1, with a smooth boundary $\partial
\Omega $ (see Fig. 1). Denote by $S_{ x_{1}} $ the cross-section
of $\Omega$ by the plane $\mathbb{R}^{d-1}$ orthogonal to the
$x_1$-axis at the point $x_{1}$ We assume that $S_{ x_{1}} $ is
bounded (and periodic with respect to~$x_{1}$).

In the simplest case, the boundary of the cross-section is a
function of the angular variables. For instance, $S_{ x_{1}} $ is
the ball of the radius $R\left( x_{1}\right)$ if $\partial \Omega
$ is the surface of revolution. However, in general it may have a
very complicated form. For example, $\Omega$ in Fig.1 contains
``fingers" (``dead ends" in the terminology of [7], [8]), and  $
S_{ x_{1}}  $ is not connected when $a<x_1<b$.

Consider the cell of periodicity in $\Omega$ defined as
$\Omega'=\Omega\bigcap\{0<x_1<1\}$. It is assumed that the compact
manifold obtained from $\Omega'$ by gluing $S_0$ to $S_1$ is
connected. For transparency, one can assume that $\Omega'$ itself
is connected. For example, $\Omega'$ in Fig.1 is connected if the
origin is slightly to the left of $a$. It is not connected if the
origin is the midpoint between $a$ and $b$, however, it still
forms a connected manifold after
gluing $S_0$ to $S_1$. One also can define a connected cell of
periodicity $\Omega'$ in Fig.1 using planes $\mathbb{R}^{d-1}$
through ${(a+b)}/{2}$ and ${(a+b)}/{2}+1$ by cutting only the
central narrow part of $\Omega$, but not cutting the fingers. In
the latter case, we use the notation $S_0,S_1$ not for the whole
cross-sections of $\Omega$, but for the parts of the boundary of
$\Omega'$ that belong to the corresponding $\mathbb{R}^{d-1}$
planes.

Denote the exterior normal to $\partial\Omega$ by $n$. Let
$x=(s,z)$, where $s=x_1,~z=(x_2,...,x_d)$. Consider the following
parabolic problem in the tube $\Omega$:
\begin{equation}\label{h1}
\frac{\partial u}{\partial t}=
\Delta u+V\frac{\partial u}{\partial s}, ~x\in
\Omega;~~\frac{\partial u}{\partial n}|_{\partial \Omega}=
0;~~u(0,x)=\varphi(\varepsilon s,z), ~x\in \Omega,
\end{equation}
where $\varphi \in L_{1,\rm{com}}(\mathbb{R}^d)$ (the space of
integrable functions with compact supports).

The main results of the first part of the paper are stated in the
following two theorems. Theorem \ref{t1} specifies the asymptotic
behavior, as $\varepsilon\downarrow 0$, of the solution $u$ of the
problem (\ref{h1}) in terms of the solution $w$ of the
one-dimensional problem
\begin{equation}\label{h2}
\frac{\partial w}{\partial t}= \sigma^2
\frac{\partial^2w}{\partial s^2}+V_{\rm eff} \frac{\partial
w}{\partial s}, ~s\in \mathbb{R};~~w|_{t=0}= w_0(s,\varepsilon s).
\end{equation}
Here $\sigma^2$ is the effective diffusivity,  $V_{\rm eff}$ is
the effective drift, and the initial function $w_0(s,\varepsilon
s)$ {\bf Remove the arguments?} is an average of $\varphi$ in
$z$-variable and will be defined later in (\ref{indata}). This
theorem provides also the main term of asymptotics of $u$ in terms
of the solution $\overline{w}$ of a one-dimensional,
$\varepsilon$-independent problem if  $\varphi$ is smooth. The
second theorem describes the effective parameters $\sigma^2$ and
$V_{\rm eff}$  in terms of the problem on the cell $\Omega'$.

We will use notation $O(\varepsilon)$ for functions $f$ such that
$|f|<C\varepsilon, ~\varepsilon\downarrow 0$.
\begin{theorem}\label{t1}
1) Let $\varphi\in L_{1,\rm{com}}(\mathbb{R}^d)$. Then for each
$t_0>0$, $$ u=w+O(\varepsilon),~ t>\frac{t_0}{\varepsilon^2}, $$
where $w$ is the solution of (\ref{h2}) and $w_0$ is defined in
(\ref{indata}).

2) Let $\varphi$ and its derivative $ \varphi_s$ be continuous
bounded functions with compact supports. Let
$t=\frac{\tau}{\varepsilon^2},~s=\frac{\varsigma-V_{\rm eff}
\tau}{\varepsilon}$ and let
$\overline{u}(\tau,\varsigma,z)=u(t,s,z)$ be the function $u$
written in the new variables. Then for each $\tau_0>0$,
\[\overline{u}=\overline{w}+O(\varepsilon),~ \tau>\tau_0,
\]
 where $\overline{w}$ is the solution of the following $\varepsilon$-independent problem
\[
\frac{\partial \overline{w}}{\partial \tau}=\sigma^2
\frac{\partial^2\overline{w}}{\partial \varsigma^2}, ~\varsigma\in
\mathbb{R};~~\overline{w}|_{t=0}=\overline{\varphi}(\varsigma)=\int_0^1w_0(s,\varsigma)d
s,
\]
\end{theorem}
\noindent \textbf{Remarks.} 1. An explicit form of
$\overline{\varphi}$ is given below in (\ref{indata}).

2. The proof of the theorem allows one to obtain explicit
estimates of the reminder terms in the formulas above through the
corresponding norms of $\varphi$ .
\\

We need to state one important lemma before we can formulate the
second theorem.  Denote by $S^l$ the lateral part of the boundary
$\partial \Omega'$ of $\Omega'$, i.e., $S^l=\partial
\Omega'\backslash [S_0\bigcup S_1]$. Consider the following
eigenvalue problem on the cell $\Omega'$:
\begin{equation}\label{hc1}
\Delta v+V\frac{\partial v}{\partial s}
=\lambda v, ~x\in \Omega';~~\frac{\partial v}{\partial n}|_{ S^l}
=0;~~v|_{S_1}=e^{i\theta}v|_{S_0},~v'_s|_{S_1}=e^{i\theta}v'_s|_{S_0}.
\end{equation}

This is an elliptic problem in a bounded domain, and for each
$\theta$ the spectrum of this problem consists of a discrete set
of eigenvalues $\lambda=\lambda_j(\theta),~j\geq 0,$ of finite
multiplicity.
\begin{lemma}\label{ml}
Let $\theta\in [-\pi,\pi]$. There exists a simple eigenvalue
$\lambda=\lambda_0(\theta)$ of problem (\ref{hc1}) in a
neighborhood $|\theta|< \delta$ of the origin $\theta=0$ and real
constants $V_{\rm eff},\sigma$ such that

1) $\lambda_0=iV_{\rm eff}\theta$$ -\sigma^2\theta^2+O(\theta^3),
\quad \theta\to 0, \quad $ where $ ~~\sigma^2> 0,$

2) $\rm{Re}\lambda_0(\theta)<0, \quad 0<|\theta|\leq \delta,$

3) $\rm{Re}$$\lambda_j(\theta)<-\gamma_1, \quad |\theta|\leq
\delta, \quad j>0, \quad \gamma_1=\gamma_1(\delta)>0,$

4) $\rm{Re}$$\lambda_j(\theta)<-\gamma_2, \quad
\delta\leq|\theta|<\pi, \quad j\geq0,\quad
\gamma_2=\gamma_2(\delta)>0.$
\end{lemma}

\begin{theorem}\label{t2}
The effective diffusivity $\sigma^2$ and effective drift $V_{\rm
eff}$ defined in (\ref{h2}) coincide with the constants introduced
in Lemma \ref{ml}.
\end{theorem}
Let us describe the averaged initial function
$\overline{\varphi}$. We need the problem adjoint to problem
(\ref{hc1}). This problem has the form:
\begin{equation}\label{hca1}
\Delta v-V\frac{\partial v}{\partial s}= \lambda v, ~x\in
\Omega';~~(\frac{\partial }{\partial n}-Vn_1)v|_{
S^l}=0;~~v|_{S_1}=e^{i\theta}v|_{S_0},~v'_s|_{S_1}=e^{i\theta}v'_s|_{S_0},
\end{equation}
where $n_1=n_1(x)$ is the first component of the normal vector
$n$. Then $\lambda=\overline{\lambda}_0(\theta)$ is an eigenvalue
of problem (\ref{hca1}). Let $v=\psi_0^*(\theta,x)$ be the
corresponding eigenfunction, and let $\pi(x)=\psi_0^*(0,x)$ be the
eigenfunction of  (\ref{hca1}) when $\theta=0$ (and
$\lambda_0=\lambda_0(0)=0$). Then $w_0, \overline{\varphi}$ are
defined as follows:
\begin{equation}\label{indata}
w_0(s,\varepsilon s)=\int_{S_{s}}\pi(s,z)
\varphi(\varepsilon s,z)dz,~~\overline{\varphi}=\int_0^1\int_{S_{s}}\pi(s,z)
\varphi(\varsigma,z)dzds,
\end{equation}
where $S_{a}$ is the cross-section of $\Omega$ by the plane through $s=a$.

The following lemma will be needed in order to prove the theorems
above. Consider the non-homogeneous problem (\ref{hc1}) in the
Sobolev space $H^2(S)$:
\begin{equation}\label{nhc1}
(\Delta +V\frac{\partial }{\partial s}-\lambda) v=f, ~x\in
\Omega';~~\frac{\partial v}{\partial n}|_{
S^l}=0;~~v|_{S_1}=e^{i\theta}v|_{S_0},~v'_s|_{S_1}=e^{i\theta}v'_s|_{S_0}.
\end{equation}
This is an elliptic problem, and the resolvent $$
R_\lambda^\theta= (\Delta +V\frac{\partial }{\partial
s}-\lambda)^{-1}:L_2(\Omega')\to H^2(\Omega') $$ is a meromorphic
in $\lambda$ operator with poles at a discrete set of eigenvalues
$\lambda=\lambda_j(\theta)$. Denote by $Q_\alpha$ the sector in
the complex $\lambda$-plane that does not contain the negative
semi-axis and is defined by inequalities
$-\pi+\alpha\leq\textrm{arg}\lambda\leq\pi-\alpha$.
\begin{lemma}\label{l0}
For every $\alpha>0$ there exist constants $R=R(\alpha)$ and
$C=C(\alpha)$ such that the region
$Q_{\alpha,R}=Q_\alpha\bigcap\{|\lambda|>R\}$ does not contain
eigenvalues $\lambda_j(\theta)$, and the following estimate is
valid for the solution $v=R_\lambda^\theta f$ of the problem
(\ref{nhc1}):
\begin{equation}\label{pell}
|\lambda|\|v\|_{L_2(\Omega')}\leq C(\alpha)\|f\|_{L_2(\Omega')}, \quad \lambda\in Q_{\alpha,R}.
\end{equation}
\end{lemma}
\noindent \textbf{Proof.} This lemma can be proved by referencing
the standard a priori estimates for parameter-elliptic problems
\cite{Grubb}. We will provide an independent proof. We multiply
equation (\ref{nhc1}) by $\overline{v}$ and integrate over
$\Omega'$. This leads to
\begin{equation}\label{gr}
-\int_{\Omega'} (|\nabla v|^2+\lambda|v|^2)dx+\int_{\Omega'}
Vv'_s\overline{v}dx=\int_{\Omega'}  f\overline{v}dx.
\end{equation}
The terms on the left in (\ref{gr}) can be estimated as follows
\[
|\int_{\Omega'}  (|\nabla v|^2+\lambda|v|^2)dx|\geq
c_1(\alpha)(\|\nabla v\|_{L_2(\Omega')}^2+|\lambda|\|v\|_{L_2(\Omega')}^2), \quad \lambda\in Q_{\alpha},
\]
and
\[
|\int_{\Omega'}  Vv'_s\overline{v}dx|\leq
\frac{c_1(\alpha)}{2}\| v'_s\|_{L_2(\Omega')}^2+\frac{V^2}{2c_1(\alpha)}\|v\|_{L_2(\Omega')}^2.
\]
We put $R=\frac{V^2}{[c_1(\alpha)]^2}$. Then the absolute value of
the left-hand side in (\ref{gr}) is estimated from below by
$\frac{c_1(\alpha)}{2}(\|\nabla
v\|_{L_2(\Omega')}^2+|\lambda|\|v\|_{L_2(\Omega')}^2)$ when
$\lambda\in Q_{\alpha,R}$, and (\ref{gr}) implies that
\begin{equation}\label{11}
\frac{c_1(\alpha)}{2}(\|\nabla v\|_{L_2(\Omega')}^2+|\lambda|\|v\|_{L_2(\Omega')}^2)
\leq \|f\|_{L_2(\Omega')}\|v\|_{L_2(\Omega')}, \quad \lambda\in Q_{\alpha,R}.
\end{equation}
We omit the first term on the left and obtain (\ref{pell}). \qed
\\

\noindent \textbf{Proof of  Lemma \ref{ml}.} Denote by
$H_\theta=\Delta+V\frac{\partial}{\partial s}$ the operator in
$L_2(\Omega')$ defined on the functions from the Sobolev space
$H^2(\Omega')$ satisfying the boundary conditions (\ref{hc1}). One
can define the parabolic semi-group $e^{tH_\theta}$. Its integral
kernel is the fundamental solution of the corresponding parabolic
boundary value problem in $\Omega'$. When $\theta=0$, this kernel
is real and positive. Thus Perron-Frobenius theorem is applicable
to the operator $e^{tH_0}$, i.e., the operator $e^{tH_0}$ has a
unique maximal eigenvalue $\mu$ such that $\mu>0$, $\mu$ is
simple, the corresponding eigenfunction is positive, all the other
eigenvalues $\mu_j$ (perhaps complex) are strictly less than $\mu$
in absolute value ($|\mu_j|<\mu$), and the operator does not have
strictly positive eigenfunctions with eigenvalues different from
$\mu$.

We note that $v_0\equiv 1$ is an eigenfunction of $H_0$ with the
eigenvalue $\lambda_0=0$. Thus $v_0$ is an eigenfunction of
$e^{tH_0}$ with the eigenvalue $\mu=1$. Since $v_0>0$, from the
Perron-Frobenius theorem it follows that $\mu=1$ is the maximal
eigenvalue of  $e^{tH_0}$. This implies that
Re$\lambda_j<\lambda_0=0$ for all the eigenvalues $\lambda_j\neq
0$ of the operator $H_0$. Since the eigenvalues  $\lambda_j$ form
a discrete set, from Lemma \ref{l0} it follows that there exists a
$\gamma>0$ such that Re$\lambda_j<-\gamma$ for all $\lambda_j\neq
0$.

Since $H_\theta$ depends analytically on $\theta$, the location of
the eigenvalues $\lambda_j$ when $\theta=0$ and Lemma \ref{l0}
imply that there exists $\delta>0$ such that the operator
$H_\theta$ has the following structure of the spectrum when
$|\theta|\leq \delta$: the operator has a simple, analytic in
$\theta$ eigenvalue $\lambda=\lambda_0(\theta),~\lambda_0(0)=0,$
and Re$\lambda_j(\theta)<-\gamma_1=-\gamma/2$ for all other
eigenvalues $\lambda_j(\theta)$ of $H_\theta$. The statement 3 of
Lemma \ref{ml} is proved. Consider now the eigenvalue
$\lambda=\lambda_0(\theta)$ for purely imaginary $\theta=iz,~z>0$.
If $v$ belongs to the domain of the operator  $H_{iz}$, then
$\overline{v}$ also belongs to the domain of the operator
$H_{iz}$. From here it follows that both $\lambda=\lambda_0(iz)$
and $\lambda=\overline{\lambda_0(iz)}$ are eigenvalues of
$H_{iz}$. Since, $\lambda=\lambda_0(\theta)$ is the unique
eigenvalue in a neighborhood of the point $\lambda=0$, it follows
that $\lambda_0(iz)$ is real. This implies statement 1 of Lemma
\ref{ml}, except the inequality $\sigma^2 > 0$. The latter
inequality will be proved below (in Theorem \ref{gt}) by a direct calculation of $\sigma^2$ (see formula (\ref{visz})). Thus it remains to justify statements 2 and 4. This will be done if we prove
that for every $\delta_1>0,$ there exists
$\gamma=\gamma(\delta_1)>0$ such that
Re$\lambda_j(\theta)<-\gamma$ for all the eigenvalues of the
operator $H_\theta$ when $\theta$ is real,
$\delta_1\leq|\theta|\leq\pi$.

Let us prove the latter estimate for the eigenvalues
$\lambda_j(\theta),~\delta_1\leq|\theta|\leq\pi$. We fix an
arbitrary rational $\theta=\theta'=\frac{\pm m}{n}$ in the set
$\delta_1\leq|\theta|\leq\pi$. Consider the domain
$\widehat{\Omega}=\Omega\bigcap\{0<x_1<n\}$, which consists of $n$
elementary cells of periodicity
$\Omega^{(j)}=\Omega\bigcap\{j<x_1<j+1\}, ~0\leq j\leq n-1
~(\Omega^{(0)}$ coincides with the previously introduced cell
$\Omega')$. The lateral side of the boundary of the domain
$\widehat{\Omega}$ will be denoted by $ \widehat{S}^l$, and the
parts of the boundary located in the planes through the points
$x_1=0$ and $x_1=n$ will be denoted by $S_0$ and $S_n$,
respectively. Let $H^{(n)}_\theta=\Delta+V\frac{\partial}{\partial
s}$ be the operator in $L_2(\widehat{\Omega})$ that corresponds to
the following analogue of the problem (\ref{hc1}):
\begin{equation}\label{hcn}
\Delta v+V\frac{\partial v}{\partial s}=
\alpha v, ~x\in \widehat{\Omega};~~\frac{\partial v}{\partial n}|_{ \widehat{S}^l}
=0;~~v|_{S_n}=e^{in\theta}v|_{S_0},~v'_s|_{S_n}=e^{in\theta}v'_s|_{S_0}.
\end{equation}
We denote its eigenvalues by $\alpha=\alpha_j$, and we keep the
notation $\lambda=\lambda_j$ for the eigenvalues of this operator
when $n=1$ (and $\widehat{\Omega}=\Omega'$).

When $\theta=\theta'$, the conditions on $S_0,S_n$ become the
periodicity condition. Thus the spectrum of $H^{(n)}_{\theta'} $
has the same structure as the spectrum of $H_0$, i.e.,
$\alpha_0=0$ is an eigenvalue with the eigenfunction $v_0=1$, and
\begin{equation}\label{mu}
\textrm{Re}\alpha_j<-\gamma<0
\end{equation}
for all other eigenvalues $\alpha_j$ of $H^{(n)}_{\theta'}$.

We compare the set $\{\lambda_j\}$ of the eigenvalues of the
operator  $H_{\theta'}$ and the set $\{\alpha_j\}$ of the
eigenvalues of the operator $H_{\theta'}^{(n)}$. The following
inclusion holds: $\{\lambda_j\}\subset\{\alpha_j\}$. Indeed, if
$v$ is an eigenfunction of the operator $H_{\theta'}$, then one
can construct the corresponding eigenfunction of
$H_{\theta'}^{(n)}$ with the same eigenvalue by defining  it as
$e^{ij\theta} v$ in each elementary cell $\Omega^{(j)}\subset
\widehat{\Omega},~0\leq j\leq n-1$. However, these two sets of
eigenvalues do not coincide. In particular, $\lambda=0$ is not an
eigenvalue of $H_{\theta'}$ (while $\alpha=0$ is an eigenvalue of
$H_{\theta'}^{(n)}$). Indeed, from the simplicity of the
eigenvalue $\alpha=0$ it follows that $\lambda=0$ could be an
eigenvalue of $H_{\theta'}$ only if $v_0=1$ is its eigenfunction,
but $v_0$ does not satisfy the boundary conditions (\ref{hc1}).
Thus, (\ref{mu}) implies that Re$\lambda_j<-\gamma<0$ for the set
of eigenvalues of the operator $H_{\theta'}$. Then the same
estimate with $\gamma/2$ instead of $\gamma$ is valid for the
eigenvalues of $H_{\theta}$ when $\theta $ is in a small enough
neighborhood of $\theta'=\frac{\pm m}{n}$. One can find a finite
covering of the set $\{\theta:\delta_1\leq|\theta|\leq\pi\}$ by
some of these neighborhoods. Thus, the desirable estimate of
eigenvalues $\lambda_j(\theta)$ is valid for all $\theta$ of the
set $\{\theta:\delta_1\leq|\theta|\leq\pi\}$.

\qed
\\

\noindent \textbf{Proof of Theorems \ref{t1} and \ref{t2}.} The
solution $u$ of problem (\ref{h1}) can be found using the Laplace
transform in $t$:
\[
u=\frac{1}{2\pi i}\int_{a-i\infty}^{a+i\infty}v(\lambda,x)e^{\lambda t}d\lambda, \quad a>0,
\]
where $v\in L_2(\Omega)$ is the solution of the corresponding
stationary problem:
\begin{equation}\label{hc4}
\Delta v+V\frac{\partial v}{\partial s}-\lambda v= -\varphi, ~x\in
\Omega;~~\frac{\partial v}{\partial n}|_{\partial \Omega}=0.
\end{equation}

In order to solve (\ref{hc4}), we apply the Bloch transform in the
variable $s=x_1$:
\[
w(x)\to \widehat{w}(\theta,x)=
\sum_{-\infty}^\infty w(s-n,z)e^{in\theta}, \quad \theta\in(-\pi,\pi), \quad x\in \Omega.
\]
This map is unitary (up to the factor $1/\sqrt{2\pi}$) from
$L_2(\Omega)$ to $L_2([-\pi,\pi])\times L_2(\Omega')$, and the
inverse transform is given by
\[
w(x)=\frac{1}{2\pi}\int_{-\pi}^\pi \widehat{w}(\theta,x)d\theta, \quad x\in \Omega.
\]
Note that function $e^{-i\theta s}\widehat{w}(\theta,x)$ is
periodic in $s$. Thus the knowledge of $\widehat{w}$ on $\Omega'$
allows one to recover $\widehat{w}$ on the whole tube $\Omega$.

The Bloch transform reduces problem (\ref{hc4}) to a problem for
$\widehat{v}$ on the cell of periodicity $\Omega'$. The latter
problem has form (\ref{nhc1}) with $f=-\widehat{\varphi}$, i.e.,
$\widehat{v}=-R_\lambda^\theta \widehat{\varphi}$. Thus
\begin{equation}\label{hc5}
u(t,x)=\frac{-1}{4\pi^2i}
\int_{a-i\infty}^{a+i\infty}\int_{-\pi}^\pi
 R_\lambda^\theta \widehat{\varphi}e^{\lambda t}d\theta d\lambda,\quad x\in \Omega.
\end{equation}
Function $R_\lambda^\theta \widehat{\varphi}$ here is defined in
$\Omega'$, but it is extended to the whole tube $\Omega $ in such
a way that $e^{-i\theta s}R_\lambda^\theta \widehat{\varphi}$ is
periodic in $s$.

Now that formula (\ref{hc5}) for $u$ is established, we are going
to study the asymptotic behavior of $u$ as $\varepsilon\to 0$. We
split $u$ as $u=u_1+u_2$, where the terms $u_1,u_2$ are given by
(\ref{hc5}) with integration in $\theta$ over sets
$|\theta|\leq\delta$ and $\delta\leq |\theta|\leq \pi$,
respectively. We choose $\delta$ small enough, so that Lemma
\ref{ml} holds, and then make it even smaller (if needed) to
guarantee that $-\frac{\gamma_1}{2}<$Re$\lambda_0(\theta)\leq
0,~|\theta|\leq \delta$.

\begin{figure}[tbph]
 \begin{center}
\includegraphics[width=0.3\columnwidth]{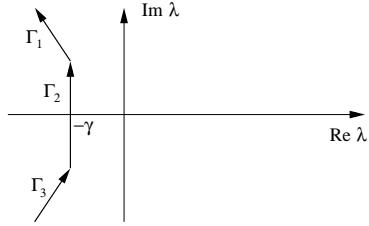}
\end{center}
\caption{Contour  $\Gamma=\Gamma_1+\Gamma_2+\Gamma_3$, where
$\gamma$ coincides with $\gamma_1$ or $\gamma_2$ defined in Lemma
\ref{ml}, and $\Gamma_2 $ is long enough, so that the estimate
(\ref{pell}) is valid on $\Gamma_1$ and $\Gamma_3$.} \label{extj3}
\end{figure}

Let us show that $u_2$ does not contribute to the main term of
asymptotics of $u$. Lemmas \ref{l0} and \ref{ml} allow us to
rewrite $u_2$ in the form
\begin{equation}\label{u2}
u_2(t,x)=\frac{-1}{4\pi^2i}\int_{\Gamma}\int_{\delta<|\theta|<\pi}
R_\lambda^\theta \widehat{\varphi}e^{\lambda t} d\theta d\lambda,\quad x\in \Omega,
\end{equation}
where $\Gamma$ is the contour shown in Fig. 2 with
$\gamma=\gamma_2$. From Lemmas \ref{l0} and \ref{ml} it follows
that
\[
\| R_\lambda^\theta \widehat{\varphi}\|_{L_2(\Omega')}\leq
 \frac{c_1}{1+|\lambda|}\|  \widehat{\varphi}\|_{L_2(\Omega')},
 \quad \lambda\in \Gamma, \quad \delta<|\theta|<\pi.
\]
Thus
\begin{eqnarray}
\int_{\delta<|\theta|<\pi}\| R_\lambda^\theta
\widehat{\varphi}\|_{L_2(\Omega')}d\theta\leq \frac{c_1}{1+|\lambda|}\|
 \widehat{\varphi}\|_{L_2(\Omega')
 \times L_2(|\theta|<\pi)}=\frac{c_2}{1+|\lambda|}\|  \varphi(\varepsilon s,z)\|_{L_2(\Omega)} \nonumber \\
\leq \frac{c_2}{(1+|\lambda|)\sqrt \varepsilon} \|
 \varphi(x)\|_{L_2(\mathbb{R}^d)}, \quad \lambda\in \Gamma. \label{123}
\end{eqnarray}
One can replace $\Omega'$ here by a shifted cell of periodicity
$\Omega^{(j)}$ defined by inequalities $j<x_1<j+1$, since the
function $e^{-i\theta x_1}R_\lambda^\theta \widehat{\varphi}$ is
periodic in $x_1$. We put (\ref{123}) with $\Omega'$ replaced by
$\Omega^{(j)}$ into (\ref{u2}) and estimate the integral over
$\Gamma$. This gives
\[
\| u_2(t,x)\|_{L_2(\Omega^{(j)})}\leq \frac{c}{\sqrt
\varepsilon}e^{-\gamma t} \| \varphi(x)\|_{L_2(\mathbb{R}^d)},
\]
where $c$ does not depend on $j$.

Now one can obtain a uniform estimate of $u_2$ by using the
following inequality for the Green function $G(t,x,y)$ of the
problem (\ref{h1}): $|G(1,x,y)|< ae^{-b(x_1-y_1)^2}$, which
implies that
\begin{eqnarray*} \nonumber
|u_2(t,x)|\leq\int_{\Omega}|G(1,x,y)u_2(t-1,y)| dy\leq
\frac{A}{\sqrt \varepsilon}e^{-\gamma t} \sum_{j=-\infty}^\infty
e^{-b(x_1-j)^2} \| \varphi(x)\|_{L_2(\mathbb{R}^d)}\\
\leq\frac{C}{\sqrt \varepsilon}e^{-\gamma t}  \|
\varphi(x)\|_{L_2(\mathbb{R}^d)}, \quad x\in \Omega,~t>1.
\label{u2b}
\end{eqnarray*}
In particular, $u_2=O(\varepsilon)$ when
$t>{t_0}/{\varepsilon^2}$. Hence $u_2$ does not contribute to the
main term of the asymptotics of $u$.

In order to study $u_1$, we also shift the contour of integration
(in $\lambda$) to $\Gamma$. Now we take $\gamma=\gamma_1$, and in
this case the simple pole of $R_\lambda^\theta$ at
$\lambda=\lambda_0(\theta)$ must be taken into account. The
residue at this pole of the integral kernel of the operator
$R_\lambda^\theta$ is equal to
$-\psi(\theta,x)\overline{\psi^*}(\theta,y)$, where $\psi$ is the
eigenfunction of the problem (\ref{hc1}) and $\psi^*$ is the
eigenfunction of the problem (\ref{hca1}) (the residue is a
negative operator). Thus
\begin{equation}\label{ab}
u_1(t,x)=\frac{1}{2\pi}\int_{|\theta|<\delta}
\int_{\Omega'}\psi(\theta,x)\overline{\psi^*}(\theta,y)
e^{\lambda_0(\theta) t}\widehat{\varphi}dyd\theta-
\frac{1}{4\pi^2i}\int_{\Gamma}\int_{|\theta|<\delta}
R_\lambda^\theta \widehat{\varphi}e^{\lambda t}d\lambda,\quad x\in \Omega,
\end{equation}
where the functions $\psi,\psi^*$ are extended to $\Omega$ using
the Bloch periodicity condition, i.e.,
\begin{equation}\label{00}
\psi(\theta,x)=e^{i\theta s}\psi_0(\theta,x),\quad \psi^*(\theta,x)=e^{i\theta s}\psi^*_0(\theta,x),
\end{equation}
where the functions $\psi_0, \psi_0^*$ are periodic with respect
to $s$. The second term in (\ref{ab}) can be estimated similarly
to $u_2$. Thus
\[
u(t,x)=\frac{1}{2\pi}\int_{|\theta|<\delta}\int_{\Omega'}\psi(\theta,x)
\overline{\psi^*}(\theta,y)e^{\lambda_0(\theta) t}\widehat{\varphi}dyd\theta+w(t,x),
\]
where $x\in \Omega,~t>1$, and
\begin{equation}\label{up}
|w|\leq\frac{C}{\sqrt \varepsilon}e^{-\gamma t}  \|
\varphi(x)\|_{L_2(\mathbb{R}^d)}=O(\varepsilon) \quad \text{if}
~~t>\frac{t_0}{\varepsilon^2},~\varphi\in L_{1,\textrm{com}}.
\end{equation}

Let $y=(s',z').$ Then
\[
\widehat{\varphi}=\sum_{-\infty}^\infty \varphi(\varepsilon(s'-n),z')e^{in\theta}.
\]
Hence (with (\ref{00}) taken into account), we have
\begin{eqnarray}
u(t,x)=\frac{1}{2\pi}\sum_{n=-\infty}^\infty \int_{|\theta|<\delta}
\int_{\Omega'}e^{i\theta(s-s'+n)}\psi_0(\theta,x)\overline{\psi_0^*}(\theta,y)
e^{\lambda_0(\theta) t}\varphi(\varepsilon(s'-n),z')dyd\theta+O(\varepsilon)\nonumber \\
\nonumber
=\frac{1}{2\pi}\sum_{n=-\infty}^\infty \int_{|\theta|<\delta}
\int_{\Omega^{(n)}}e^{i\theta(s-s')}\psi_0(\theta,x)\overline{\psi_0^*}
(\theta,y)e^{\lambda_0(\theta) t}\varphi(\varepsilon s',z')dyd\theta+O(\varepsilon)  \\ \label{21}
=\frac{1}{2\pi}\int_{|\theta|<\delta}\int_{\Omega}e^{i\theta(s-s')}
\psi_0(\theta,x)\overline{\psi_0^*}(\theta,y)e^{\lambda_0(\theta) t}
\varphi(\varepsilon s',z')dyd\theta+O(\varepsilon).
\end{eqnarray}

We split the double integral $I$ above in two parts $I=I_1+I_2$,
where the integration in $\theta$ in the part $I_1$ extends only
over the segment $|\theta^3t|<1, ~t\geq \delta^{-3},$ and
$I_2=I-I_1$. From Lemma \ref{ml} it follows that
Re$\lambda_0(\theta)<-\frac{\sigma^2}{2}\theta^2,
~|\theta|<\delta,$ if $\delta$ is small enough.  Since we can take
$\delta$ as small as we please, and the functions
$\psi_0,\psi_0^*$ are bounded, it follows that
\[
I_2\leq
C\int_{|\theta|>t^{-1/3}}\int_{\Omega}e^{-\frac{\sigma^2}{2}\theta^2t}|\varphi(\varepsilon
s',z')|dyd\theta =\frac{\|
\varphi(x)\|_{L_1(\mathbb{R}^d)}}{\varepsilon}O(e^{-\frac{\sigma^2}{2}t^{1/3}})=O(\varepsilon),
\quad t>\frac{\tau_0}{\varepsilon^2}.
\]
We replace $\lambda_0(\theta)$ in the integral $I_1$ by its quadratic approximation found in Lemma \ref{ml}:
\[
e^{\lambda_0(\theta) t}=e^{(iV_{\rm eff}\theta-\sigma^2\theta^2)
t}(1+O(|\theta|^3t)).
 \]
If only the remainder term is taken into account, then the corresponding integral can be estimated by
\begin{eqnarray*}
C\int_{|\theta|<t^{-1/3}}\int_{\Omega}|\theta|^3te^{-\sigma^2\theta^2t}\varphi(\varepsilon
s',z')dyd\theta\leq
\frac{C}{\varepsilon}\|\varphi(y)\|_{L_1(\mathbb{R}^d)}
\int_{-\infty}^\infty|\theta|^3te^{-\sigma^2\theta^2t}d\theta\\
=\frac{C}{t\varepsilon}\|\varphi(y)\|_{L_1(\mathbb{R}^d)}=O(\varepsilon),
\quad \textrm{if} ~t>\frac{\tau_0}{\varepsilon^2}.
\end{eqnarray*}
From here and (\ref{21}) it follows that
\[
u(t,x)=\frac{1}{2\pi}\int_{|\theta|<t^{-1/3}}\int_{\Omega}e^{i\theta(s-s')}\psi_0(\theta,x)
\overline{\psi_0^*}(\theta,y) e^{(iV_{\rm
eff}\theta-\sigma^2\theta^2) t}\varphi(\varepsilon
s',z')dyd\theta+O(\varepsilon), \quad
t>\frac{\tau_0}{\varepsilon^2}.
\]

We put $\theta=0$ in the arguments of functions $\psi_0$ and
$\overline{\psi_0^*}$, Since these functions are periodic,
$\psi_0=1$ and $\psi_0^*(0,y)=\pi(y)$ is real, we obtain that
$\psi_0(\theta,x)\overline{\psi_0^*}(\theta,y)=\pi(y)+O(|\theta|)$.
We put this relation into the formula above and note that the
integral with the term $O(|\theta|)$ does not exceed
\begin{eqnarray*}
C\int_{|\theta|<t^{-1/3}}\int_{\Omega}|\theta|e^{-\sigma^2\theta^2t}\varphi(\varepsilon
s',z')dyd\theta \leq
\frac{C}{\varepsilon}\|\varphi(y)\|_{L_1(\mathbb{R}^d)}
\int_{-\infty}^\infty|\theta|e^{-\sigma^2\theta^2t}d\theta\\
=\frac{C}{t\varepsilon}\|\varphi(y)\|_{L_1(\mathbb{R}^d)}=O(\varepsilon),
\quad \textrm{if} ~t>\frac{\tau_0}{\varepsilon^2}.
\end{eqnarray*}
Hence if $t>{\tau_0}/{\varepsilon^2}$, then
\[
u(t,x)=\frac{1}{2\pi}\int_{|\theta|<t^{-1/3}}\int_{\Omega}e^{i\theta(s-s')}\pi(y)
e^{(iV_{\rm eff}\theta-\sigma^2\theta^2) t}\varphi(\varepsilon
s',z')dyd\theta+O(\varepsilon) , \quad y=(s',z').
\]
We can replace here the integration in $\theta$ over the interval
$|\theta|<t^{-1/3}$ by the integration over the whole line since
the difference between the corresponding integrals decays
exponentially as $t\to\infty$. Thus
\[
u(t,x)=\frac{1}{2\pi}\int_{-\infty}^\infty\int_{\Omega}e^{i\theta(s-s')}\pi(y)
e^{(iV_{\rm eff}\theta-\sigma^2\theta^2) t}\varphi(\varepsilon
s',z')d y d \theta+O(\varepsilon), \quad t>\tau_0/\varepsilon^2.
\]

The integral $w=w(t,s)$ above does not depend on $z$. By simple
differentiation one can check that the integral satisfies equation
(\ref{h2}). Function $w(0,s)$ is the Fourier transform in $s'$
followed by its inverse (in $\theta$), i.e.,
\begin{equation}
w(0,s)=\frac{1}{2\pi}\int_{-\infty}^\infty\int_{-\infty}^\infty\int_{S_{s'}}e^{i\theta(s-s')}\pi(s',z')
\varphi(\varepsilon s',z')dz'ds'd\theta
=\int_{S_{s}}\pi(s,z')
\varphi(\varepsilon s,z')dz',
\end{equation}
where $S_{a}$ is the cross-section of $\Omega$ by the plane
through $s=a$. Thus $w$ is the solution of problem (\ref{h2}) with
$w_0$ given in (\ref{indata}), and the first statement of Theorem
\ref{t1} is proved.

In order to prove the second statement of Theorem \ref{t1}, we
will show that $w-\overline{w}=O(\varepsilon), \tau>\tau_0$. We
denote function $w$ in new coordinates by $\widehat{w}$, i.e.,
$\widehat{w}(\tau,\varsigma)=w(\frac{\tau}{\varepsilon^2},\frac{\varsigma-V_{\rm
eff} \tau}{\varsigma}) $. Obviously, $\widehat{w}$ satisfies the
same equation as the equation for $\overline{w}$, and
$\widehat{w}(0,\varsigma)=w_0(\frac{\varsigma}{\varepsilon},\varsigma)$.
The solution of the initial problem for $\widehat{w}$ is the
convolution of the smooth (when $\tau>\tau_0)$ kernel
$K=\frac{1}{\sqrt{4\pi\sigma\tau}}e^{\frac{-\varsigma^2}{4\sigma\tau}}$
and $w_0(\frac{\varsigma}{\varepsilon},\varsigma)$. The second
factor is periodic in the first argument (see (\ref{indata})).
Thus, the convolution differs by $O(\varepsilon)$ from the same
convolution when the second factor is replaced by its average with
respect to the first argument. This second convolution is
$\overline{w}$. This completes the proof of the second statement
of the Theorem \ref{t1}.

It remains to note that in the proof of Theorem \ref{t1} it was
shown that the effective drift and the effective diffusivity were
defined by the coefficients of the Taylor expansion of the
eigenvalue $\lambda_0(\theta)$. Thus, Theorem \ref{t2} was also
established.

\qed

In conclusion of this section we will provide some formulas that
can be useful for practical evaluation of the effective drift
$V_{\rm eff}$ and the effective diffusivity $\sigma^2$.

Let
\begin{equation}\label{ass}
v=1+i\theta v_1(x)-\theta^2 v_2(x) +O(\theta^3), \quad \theta\to 0,
\end{equation}
be the Taylor expansion at zero of the principal eigenfunction of
the problem (\ref{hc1}) with the eigenvalue
$\lambda=\lambda_0(\theta)$. We plug (\ref{ass}) and the expansion
for $\lambda_0(\theta)$ from Lemma \ref{ml} into (\ref{hc1}) and
take into account that conditions on $S_0,S_1$ in (\ref{hc1}) can
be rewritten as the periodicity in $s=x_1$ of the function
$e^{ix\theta}v$. Then we obtain the following problems for the
coefficients $v_1, v_2$:
\begin{equation}\label{vs1}
\Delta v_1+V\frac{\partial v_1}{\partial s}=V_{\rm eff},~~ ~x\in
 \Omega';~~\frac{\partial v_1}{\partial n}|_{ S^l}=0;~~v_1=s+\psi_1(x),
\end{equation}
\begin{equation}\label{vs2}
\Delta v_2+V\frac{\partial v_2}{\partial s}=V_{\rm eff}
v_1+\sigma^2,~~ ~x\in \Omega';~~\frac{\partial v_2}{\partial n}|_{
S^l}=0;~~v_2=\frac{s^2}{2}+s\psi_1(x)+\psi_2(x),
\end{equation}
where the functions $\psi_1,\psi_2$ are periodic. Function $v_1$ is called the \textit{harmonic coordinate}.  Usually the harmonic coordinate satisfies a homogeneous equation, but this equation (see (\ref{vs1})) becomes inhomogeneous in the presence of a drift in the problem.
 Denote by
$\pi=\pi(x)$ the principal eigenfunction with eigenvalue
$\lambda=0$ for the adjoint problem (\ref{hca1}) with $ \theta
=0$, i.e.,
\begin{equation}\label{hca1a}
\Delta \pi-V\frac{\partial \pi}{\partial s}= 0, ~x\in
\Omega';~~(\frac{\partial }{\partial n}-Vn_1)\pi|_{
S^l}=0;~~\pi|_{S_1}=\pi|_{S_0},~\pi'_s|_{S_1}=\pi'_s|_{S_0}.
\end{equation}
This is a positive function due to the Perron-Frobenius theorem
(see details in the proof of Lemma \ref{gt}). We  normalize $\pi$
by the condition $\int_{\Omega'}\pi dx=1$. This function is
\textit{the density of the invariant measure}. Note that $\pi
(x)\equiv 1/|\Omega'|$ if $V=0$.
\begin{theorem}\label{gt}
The effective drift $V_{\rm eff}$ can be found from either of the
following three formulas:
\begin{equation}\label{vv}
 V_{\rm eff}=\int_{S}(V\pi-\pi'_s) dS=V-\int_{\Omega'}\pi'_s dx
 =V-\int_{\partial S^l}n_1\pi dS,
\end{equation}
where $S$ is a cross-section of the domain $\Omega$ by an
arbitrary hyperplane $s={\rm{const.}}$, $n_1$ is the first
component of the outward normal vector $n$, and $S^l=\partial
\Omega'\backslash [S_0\bigcup S_1]$ is the lateral part of the boundary
of $\Omega'$. The effective
diffusivity is given by
\begin{equation}\label{visz}
\sigma^2=\int_{\Omega'}|\nabla v_1|^2\pi dx.
\end{equation}
\end{theorem}

\noindent \textbf{Remark.} If $V=0$, then formula (\ref{visz}) is equivalent to the following one
\[
\sigma^2=1-\frac{\int_{\Omega'}|\nabla \psi_1|^2dx}{|\Omega'|}.
\]
The latter formula shows that the effective diffusivity is always smaller than in the free space if $\Omega$ is not a cylinder (i.e., if $\psi_1$ is not identical zero).

\noindent \textbf{Proof.} The remark will be justified at the end of the proof of the theorem.
Let $L=\Delta+V\frac{\partial }{\partial
s}$ be the differential expression in the left-hand side of
(\ref{vs1}), and let $L^*=\Delta-V\frac{\partial }{\partial s}$ be
the conjugate expression (see (\ref{hca1a})). Since $L^*\pi=0$ and
$v_1$ satisfies (\ref{vs1}), we have
\[
\int_{\Omega'}(Lv_1)\pi dx-\int_{\Omega'}v_1 L^*\pi dx={V_{\rm
eff}}\int_{\Omega'}\pi dx={V_{\rm eff}}.
\]
The left-hand side here can be rewritten using the Green's formula
and the divergence theorem
applied to the vector field $\overrightarrow{F}=(
v_1\pi,0,0)$. This leads to
\begin{equation*}
{V_{\rm eff}}\int_{\Omega'}\pi dx= \int_{\partial\Omega'}[(\Delta
v_1)\pi -v_1\Delta \pi+V\frac{\partial (v_1\pi)}{\partial s} ]dx=
\int_{\partial\Omega'}(\frac{\partial v_1}{\partial n}\pi
-v_1\frac{\partial \pi}{\partial n}+Vv_1\pi n_1)dS.
\end{equation*}
The second integrand vanishes on $S^l$ since $\pi$ satisfies the
boundary condition (\ref{hca1a}) on $S^l$ and $\frac{\partial
v_1}{\partial n}=0$ on $S^l$. Thus
\begin{equation}\label{vvv}
{V_{\rm eff}}\int_{\Omega'}\pi dx=\int_{S_0\bigcup S_1}
(\frac{\partial v_1}{\partial n}\pi -v_1\frac{\partial
\pi}{\partial n}+Vv_1\pi n_1)dS.
\end{equation}
We substitute here $v_1=s+\psi_1$. The integral with $\psi_1$ is
zero due to the periodicity of the functions $\psi_1$ and $\pi$.
Thus (\ref{vvv}) holds with $v_1$ replaced by $s$. If we also take into account that
\[
\int_{S_0\bigcup S_1}
\frac{\partial s}{\partial n}\pi dS=0,
\]
then we obtain the first equality (\ref{vv}) with $S=S_1$. Then
this equality holds with any $S$ since $V_{\rm eff}$ is invariant
with respect to the shift $s\to s+a$. Let us provide another way
to show that the middle term in (\ref{vv}) does not depend on the
choice of $S$. Let $\Omega_{a,b}=\Omega\bigcap\{a<s<b\}$. Then
\begin{eqnarray*}
0=\int_{\Omega_{a,b}}L^*\pi dx=\int_{\partial\Omega_{a,b}}(\pi'_n -V\pi n_1)dS=
\int_{S_a\bigcup S_b}(\pi'_n -V\pi n_1)dS\\=\int_{S_b}(\pi'_s -V\pi )dS-\int_{S_a}(\pi'_s -V\pi )dS.
\end{eqnarray*}

In order to prove the second equality (\ref{vv}), we write the
first one with $S=S_a$ and integrate with respect to $a$ over the
interval $(0,1)$. The last equality (\ref{vv}) follows from the divergence theorem.

The proof of (\ref{visz}) is similar. We have
\[
\int_{\Omega'}({V_{\rm eff}}v_1+\sigma^2 )\pi
dx=\int_{\Omega'}(Lv_2)\pi
 dx-\int_{\Omega'}v_2 L^*\pi dx=\int_{S_0\bigcup S_1}(\frac{\partial v_2}{\partial n}\pi -
 v_2\frac{\partial \pi}{\partial n}+Vv_2\pi n_1)dS.
\]
We plug here
$v_2=\frac{s^2}{2}+s\psi_1+\psi_2$. The integral with $\psi_2$
vanishes due to the periodicity of $\psi_2$ and $\pi$. Hence $v_2$
can be replaced by $\frac{s^2}{2}+s\psi_1$. We reduce the integral
over $S_0$ to an integral over $S_1$ using the substitution $s\to
s-1$. Taking into account the periodicity of $\psi_1$ and $\pi$ we
obtain
\begin{eqnarray}
\int_{\Omega'}({V_{\rm eff}}v_1+\sigma^2 )\pi
dx=\int_{S_1}[(1+\frac{\partial \psi_1}{\partial s})\pi-
(\frac{1}{2}+\psi_1)\frac{\partial \pi}{\partial
s}+V(\frac{1}{2}+\psi_1)\pi]dS  \nonumber\\
=\int_{S_1}[\frac{\partial v_1}{\partial s}\pi-v_1\frac{\partial
\pi}{\partial s}+Vv_1\pi]dS- \frac{1}{2}{V_{\rm eff}}. \nonumber
\end{eqnarray}
The last relation is the consequence of the first equality
(\ref{vv}) and the equality $v_1=s+\psi_1$. Hence,
\begin{equation}\label{aq}
\sigma^2=\int_{S_0}[\frac{\partial v_1}{\partial
s}\pi-v_1\frac{\partial \pi}{\partial s}+Vv_1\pi]dS-{V_{\rm
eff}}\int_{\Omega'}v_1\pi dx- \frac{1}{2}{V_{\rm eff}}.
\end{equation}
It remains to show that
\begin{equation}\label{aq1}
\int_{\Omega'}|\nabla v_1|^2\pi dx=\int_{S_0}[\frac{\partial
v_1}{\partial s}\pi-v_1\frac{\partial \pi}{\partial
s}+Vv_1\pi]dS-{V_{\rm eff}}\int_{\Omega'}v_1\pi dx-
\frac{1}{2}{V_{\rm eff}}.
\end{equation}

We note that $L(v_1^2)=2v_1Lv_1+2|\nabla v_1|^2=2{V_{\rm
eff}}v_1+2|\nabla v_1|^2$. Thus
\begin{eqnarray*}
\int_{\Omega'}|\nabla v_1|^2\pi
dx=\frac{1}{2}\int_{\Omega'}L(v_1^2)\pi dx-{V_{\rm
eff}}\int_{\Omega'}v_1 \pi dx \\
=
\int_{S_0\bigcup S_1}[v_1\frac{\partial v_1}{\partial n}\pi-
\frac{1}{2}v_1^2(\frac{\partial \pi}{\partial n}-V\pi
n_1)]dS-{V_{\rm eff}}\int_{\Omega'}v_1 \pi dx.
\end{eqnarray*}
We took into account here that the integrand of the first term on
the right vanishes at $S^l$. The last inequality implies
(\ref{aq1}) if the integral over $S_1$ in the equality above is
rewritten in terms of the integral over $S_0$ using the first
formula (\ref{vv}) and periodicity of the functions $\pi$ and
$v_1-s$. The proof of the theorem is complete. Now let us justify
the remark.

From the divergence theorem and periodicity of $\psi_1$ it follows that
\[
\int_{\Omega'}\frac{\partial \psi_1}{\partial s}dx=\int_{\partial\Omega'}
\psi_1n_1dS=\int_{S^l} \psi_1n_1dS.
\]
Since $n_1=\frac{\partial s}{\partial n}$ and the latter function
on $S^l$ is equal to $-\frac{\partial \psi_1}{\partial n}$ (due to
(\ref{vs1})), we have
 \[
\int_{\Omega'}\frac{\partial \psi_1}{\partial s}dx=\int_{\partial\Omega'}
\psi_1n_1dS=-\int_{S^l} \psi_1\frac{\partial \psi_1}{\partial n}dS=
-\int_{\partial\Omega'} \psi_1\frac{\partial \psi_1}{\partial n}dS.
\]
Now we recall that $\Delta\psi_1=\Delta(v_1-s)=0$. Hence the
latter formula together with the Green formula imply
\begin{equation}\label{az}
\int_{\Omega'}\frac{\partial \psi_1}{\partial s}dx=
-\int_{\Omega'}|\nabla\psi_1|^2dx.
\end{equation}
Further, if $V=0$, then $\pi=1/|\Omega'|$.
 From $v_1=\psi_1+s$, (\ref{visz}) and (\ref{az}) it follows that
\begin{eqnarray*}
\sigma^2|\Omega'|=\int_{\Omega'}|\nabla v_1|^2dx=\int_{\Omega'}(1+2\frac{\partial \psi_1}{\partial s}+|\nabla\psi_1|^2)dx=
\int_{\Omega'}(1-|\nabla\psi_1|^2)dx
\end{eqnarray*}

\qed

\begin{theorem}
If $V=0$ then ${V_{\rm eff}}=0$, and ${V_{\rm
eff}}=\sigma_0^2V+O(V^2)$ as $V\to 0$ where
$\sigma_0^2=\sigma^2|_{V=0}>0$ is the effective diffusivity at
$V=0$. In particular, if $|V|$ is small enough, then ${V_{\rm
eff}}\neq 0$ when  $V\neq 0$ and ${V_{\rm eff}}$ is a monotone
function of $V$.
\end{theorem}
\noindent \textbf{Proof.} From (\ref{hca1a}) it follows that $\pi$
is an analytic function of $V$. Then (\ref{vv}) implies that
${V_{\rm eff}}$ is analytic in $V$, and (\ref{vs1}) implies that
$v_1$ and $\psi_1$ depend on $V$ analytically. We extend
$\pi=\frac{1}{|\Omega'|}+V\pi_1+O(V^2)$ in the Taylor series at
$V=0$. Then (\ref{hca1a}) leads to the following problem for
$\pi_1$:
\[
\Delta \pi_1=0, ~~x\in \Omega' ;~~\frac{\partial\pi_1}{\partial n}|_{S^l}=\frac{n_1}{|\Omega'|};~~\pi_1|_{S_1}=\pi_1|_{S_0},~(\pi_1)'_s|_{S_1}=(\pi_1)'_s|_{S_0}.
\]
Since $n_1=\frac{\partial s}{\partial n}$, only the factor
$-1/|\Omega'|$ in the boundary condition makes the problem for
$\pi_1$ different from the problem for $\psi_1^0=\psi_1|_{V=0}$.
The latter problem can be obtained from (\ref{vs1}) if we put
there $V={V_{\rm eff}}=0$. Hence
$\pi_1=-\frac{\psi_1^0}{|\Omega'|}$, i.e.,
$\pi=\frac{1}{|\Omega'|}-\frac{V\psi_1^0}{|\Omega'|}+O(V^2)$. We
put this into the second of relations (\ref{vv}) and obtain that
\[
{V_{\rm
eff}}=V(1+\frac{\int_{\Omega'}(\psi_1^0)_s'dx}{|\Omega'|})+O(V^2)=V(1-\frac{\int_{\Omega'}|\nabla\psi_1^0|^2dx}{|\Omega'|})+O(V^2).
\]
The last relation follows from (\ref{az}). It remains only to use the Remark after Theorem \ref{gt}.

\qed

\section{Periodic tubes with nearly separated dead zones. }

Consider  the problem (\ref{h1}) with $V=0$ (without drift) in a
periodic domain $\Omega=\Omega(\varepsilon) \subset
\mathbb{R}^{d}$ introduced in the previous section, that has the
form of an $\varepsilon$-independent periodic tube $\Omega_0$ with
a periodic system of cavities connected to the tube $\Omega_0$ by
narrow channels (see Fig. 3). In this section we'll assume that
$d\geq 3$. We do not impose restrictions on the shape of channels,
except the periodicity condition, smoothness of the boundary
$\partial \Omega (\varepsilon)$  and the assumption that the
channel enters the cell of periodicity $\Omega'_0$ of the main
tube in an $\varepsilon$ neighborhood of some point
$x_0\in\partial\Omega'_0$. For simplicity we will assume that
there is only one cavity and one channel on each period, but
practically no changes are needed to extend all the arguments to
the case of several cavities per period or multiple channels
connecting the cavity with the tube.

\begin{figure}[tbph]
\begin{center}
\includegraphics[width=0.4\columnwidth]{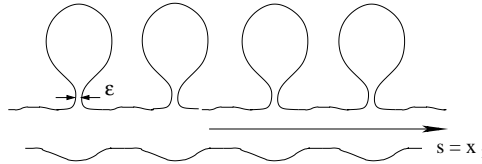}
\end{center}
\caption{Periodic tube with nearly separated cavities (dead
zones).} \label{extj1}
\end{figure}

Denote by $\Omega'(\varepsilon)=\Omega'_0\bigcup
\Omega'_1(\varepsilon)$ one cell of periodicity of the domain
$\Omega(\varepsilon)$, where $\Omega'_0$ is the cell of
periodicity of the main tube $\Omega_0$ and
$\Omega'_1(\varepsilon)$ is the union of the cavity and the
channel on this period. We assume that the cell $\Omega'_0$ is
defined by the restriction $0<s=x_1<1$. We denote by $S_0,~S_1$
the parts of the boundary of  $\Omega'_0$ that belong to the
planes through the points $s=0$ and $s=1$, respectively. Let
$S^l=\partial \Omega'_0\backslash [S_0\bigcup S_1]$ be the lateral
part of the boundary $\partial \Omega'_0$ of the cell of the main
tube.

In order to describe the effective diffusivity $\sigma^2$ for the
problem in $\Omega(\varepsilon)$ and the effective diffusivity
$\sigma_0^2$ for the problem in the main tube $\Omega_0$ (without
the cavities and channels), we introduce the following auxiliary
problem
\begin{equation}\label{axp}
\Delta u=0,~~x\in \Omega_0;~~ u=s+\psi, ~~\textrm{where}~\psi~\textrm{is 1-periodic in }s;~~u'_n|_{S^l}=0.
\end{equation}
We will call function $u$ the harmonic coordinate (for the domain
$\Omega_0$). This function was introduced in the previous section
where it was denoted by $v_1$ (see (\ref{vs1})).
\begin{theorem} Let $V=0$. Then the following formulas are valid for the effective diffusivity $\sigma_0^2$
in the case of a periodic tube $\Omega_0$ and for the effective diffusivity $\sigma_0^2(\varepsilon)$
in the case of the tube $\Omega_0$ with cavities:
\begin{equation}\label{sig0}
\sigma_0^2=\frac{\int_{\Omega'_0}|\nabla u|^2dx}{|\Omega'_0|},
\end{equation}
\begin{equation}\label{sig}
\sigma^2(\varepsilon)=\frac{|\Omega'_0|\sigma_0^2}{|\Omega'_0|+
|\Omega'_1(\varepsilon)|}+O(\varepsilon^{d-2}), \quad \varepsilon\to 0.
\end{equation}
\end{theorem}

\noindent  \textbf{Remark.} When $\Omega_0$ is a cylinder, formula
(\ref{sig}) under some assumptions on the connecting channels can
be found in \cite{bm1}, \cite{bm}.
\\

\noindent \textbf{Proof.} In the case of
domain $\Omega_0$, we will use notations $u,\psi$ for functions
$v_1,\psi_1$, defined in (\ref{ass}), (\ref{vs1}) (compare (\ref{vs1}) and (\ref{axp})), and we preserve
the original notations $v_1,\psi_1$ when the problem in
$\Omega(\varepsilon)$ is considered.

We note that $\pi$ is a constant when $V=0$. From the normalization condition ($\int_{\Omega'}\pi dx=1$) it follows that
$\pi=1/|\Omega_0'|$ for the problem in $\Omega_0$ and $\pi=1/|\Omega_0'(\varepsilon)|$ for the problem in $\Omega_0(\varepsilon)$. Hence, formula (\ref{sig0}) follows from
(\ref{visz}). In fact, (\ref{visz}) was derived without any specific
assumptions on the periodic domain $\Omega$, and it is valid for
both domains $\Omega_0$ and  $\Omega(\varepsilon)$. Thus
\begin{equation}\label{sig2}
\sigma^2(\varepsilon)=\frac{\int_{\Omega'(\varepsilon)}|\nabla v_1|^2dx}{|\Omega'(\varepsilon)|}.
\end{equation}
We are going to compare  the functions $u$ and $v_1$ and
derive (\ref{sig}) from (\ref{sig2}) and (\ref{sig0}).

Let $x_0\in S^l$ be the point on the lateral side of the cell
$\Omega_0'$ where the channel enters the main tube. Harmonic coordinates $u$ and $v_1$
(for tubes $\Omega_0$ and $\Omega(\varepsilon)$, respectively) are defined up to
arbitrary additive constants.
We fix these constants assuming that $u(x_0)=v_1(x_0)=0$.

We fix a
function $\alpha=\alpha(\varepsilon,x)\in C^\infty(\Omega_0')$
such that $\alpha=0$ when $|x-x_0|<2\varepsilon$, $\alpha=1$ when
$|x-x_0|>3\varepsilon$ and $\frac{\partial\alpha}{\partial n}=0$
on $S^l$. We consider function $w=\alpha(x)u(x)$ and
extend it by zero in the channel and the cavity. Then $w\in
C^\infty(\Omega'(\varepsilon))$.

Let us estimate $v_1-w$. Obviously, this difference satisfies the
following relations in $\Omega(\varepsilon)$:
\begin{equation}\label{v1w}
\Delta (v_1-w)=f:=2\nabla \alpha\nabla u+u\Delta  \alpha
,~~x\in\Omega(\varepsilon);  \quad \frac{\partial
(v_1-w)}{\partial n}=0,~~ x\in  S^l(\varepsilon),
\end{equation}
and it satisfies the periodicity condition on $S_0,S_1$. Function
$u$ is smooth and $\varepsilon$-independent. Function $\nabla
\alpha$ has order $O(\varepsilon^{-1})$, and its support belongs
to a ball of radius $3\varepsilon$. Thus $\|\nabla \alpha\nabla
u\|_{L_2(\Omega'(\varepsilon))}=O(\varepsilon^\frac{d-2}{2}), ~
\varepsilon\to 0$. A similar estimate is valid for
$u\Delta \alpha$ since $\Delta
\alpha=O(\varepsilon^{-2})$ and $|u|<C\varepsilon$ on the
support of $\Delta \alpha$. Hence
$\|f\|_{L_2(\Omega'(\varepsilon))}=O(\varepsilon^\frac{d-2}{2}), ~
\varepsilon\to 0.$ We also take into account that the support of $f$ belongs to $\Omega'_0$. Thus from the Green  formula it follows that
\begin{equation}\label{lk}
\|\nabla(v_1-w) \|^2_{L_2(\Omega'(\varepsilon))}\leq
 C\|f\|_{L_2(\Omega'(\varepsilon))}\|v_1-w \|_{L_2(\Omega'_0)}=O(\varepsilon^\frac{d-2}{2})\|v_1-w \|_{L_2(\Omega'_0)}, \quad \varepsilon\to 0.
\end{equation}
Since domain $\Omega'_0$ does not depend on $\varepsilon$ and $(v_1-w)(x_0)=0$, we have $\|v_1-w\|_{L_2(\Omega'_0)}\leq C\|\nabla(v_1-w)\|_{L_2(\Omega'_0)}$. Hence from (\ref{lk}) it follows that
\begin{equation}\label{lk1}
\|\nabla(v_1-w) \|_{L_2(\Omega'(\varepsilon))}= O(\varepsilon^\frac{d-2}{2}). \quad \varepsilon\to 0.
\end{equation}
Thus one can make the following changes in formula (\ref{sig2})
with the accuracy of $O(\varepsilon^{d-2})$: replace $v_1$
by $\alpha u$, replace integration over
$\Omega'(\varepsilon)$ by the integration over $\Omega'_0$, and
then drop $\alpha$. In other words, one can replace the
numerator in (\ref{sig2}) by the numerator from (\ref{sig0}) plus
$O(\varepsilon^{d-2})$. Then it remains only to use
(\ref{sig0}) and express the latter numerator through
$\sigma_0^2$. \qed


\begin{thebibliography}{99}



\bibitem{bm1}  A. Berezhkovskii, L. Dagdug,
Analytic treatment of biased diffusion in tubes with periodic dead
ends, J. Chemical Phys., 134 (2011), 124109.

\bibitem{CoVann} C. Conca, M. Vanninathan,
Homogenization of periodic structures via Bloch decomposition,
Siam J. Appl. Math. 57, No. 6 (1997) pp. 1639-1659.

\bibitem{CoOrVann} C. Conca, R. Orive, M. Vanninathan,
Bloch approximation in homogenization on bounded domains,
Asymptot. Anal. 41 (2005), No. 1, 71-91.



\bibitem{bm} L. Dagdug, A. Berezhkovskii, Y. Makhnovskii, V. Zitserman,
Transient diffusion in the tube with dead ends, J. Chemical Phys., 127 (2007), 224712.

\bibitem{Grubb}
G. Grubb, Functional Calculus of Pseudodifferential Boundary
Problems, Birkhauser, Boston, 1996.

\bibitem{rei} P. Reimann, Brownian motors:
noisy transport far from equilibrium, Physical reports, 361 (2002), 57-265.



\end{thebibliography}
\end{document}